\documentclass[11pt]{article}
\usepackage{ascmac,amsmath}
\usepackage{bm}
\usepackage{graphicx,epsfig}
\usepackage{geometry}
\makeatletter
\renewenvironment{thebibliography}[1]
{\section*{\refname\@mkboth{\refname}{\refname}}%
  \list{\@biblabel{\@arabic\c@enumiv}}%
       {\settowidth\labelwidth{\@biblabel{#1}}%
        \leftmargin\labelwidth
        \advance\leftmargin\labelsep
 \setlength\baselineskip{11pt}%
        \@openbib@code
        \usecounter{enumiv}%
        \let\p@enumiv\@empty
        \renewcommand\theenumiv{\@arabic\c@enumiv}}%
  \sloppy
  \clubpenalty4000
  \@clubpenalty\clubpenalty
  \widowpenalty4000%
  \sfcode`\.\@m}
 {\def\@noitemerr
 {\@latex@warning{Empty `thebibliography' environment}}%
\endlist}
\makeatother
\begin{document}
\centerline{{\sl Genshikaku Kenkyu Suppl.} No. 000 (2012)}
\begin{center} 
\vskip 2mm
{\Large\bf

Analysis of heavy hyperhydrogen $^6_\Lambda$H
\hspace{-1mm}\footnote{Presented at the International Workshop on Strangeness 
Nuclear Physics (SNP12), August 27 - 29, 2012, \\
\hspace*{5mm} Neyagawa, Osaka, Japan.}
}\vspace{5mm}

{
Theingi$^a$, Khin Swe Myint$^a$, and Y. Akaishi$^b$
}\bigskip

{\small
$^a$Department of Physics, Mandalay University, Mandalay, Myanmar\\ 
$^b$RIKEN, Nishina Center, Wako, Saitama 351-0198, Japan\\ 
}
\end{center}
\vspace{3mm}

\noindent
{\small \textbf{Abstract}:\quad
We have revisited the possible existence and the binding mechanism of heavy hyperhydrogen $^6_\Lambda$H. This $\Lambda$ hypernucleus is enriched in information about $\Lambda$ hypernuclear dynamics which would shed a light on significant role of coherent $\Lambda$-$\Sigma$ coupling effect in neutron-rich nuclear medium, discussed by the authors. Recently, reports on an experimental evidence of $^6_\Lambda$H with theoretical comparison have appeared, wherein the significance of the coherent coupling in $^6_\Lambda$H has been criticized. In this report, we are revising our previous work and give further theoretical analysis to confirm the influence of the $\Lambda N$-$\Sigma N$ coupling effect in the binding mechanism of $^6_\Lambda$H.
}%


\section{Introduction}

Super-heavy hydrogen $^5$H was successfully produced via $^1$H($^6$He, $^2$He)$^5$H reaction \cite{Korsheni01}. It is a resonance state at $1.7 \pm 0.3$ MeV above the $n+n+t$ threshold with a width of $1.9 \pm 0.4$ MeV. When a $\Lambda$ particle is added to it, "hyper-heavy" hydrogen $^6_\Lambda$H is obtained. Three events of the heavy hyperhydrogen $^6_\Lambda$H have been identified in the FINUDA experiment at DA$\Phi$NE, Frascati by observing $\pi^+$ mesons from the production reaction on $^6$Li target, in coincidence with $\pi^-$ mesons from its weak decay \cite{Agnello12,Agnello12a}. The $^6_\Lambda$H binding energy with respect to $^5$H$+\Lambda$ has been determined jointly from the production and the decay to be $B_\Lambda$ = $4.0 \pm 1.1$ MeV \cite{Agnello12,Agnello12a}.

\begin{figure}[htb!] 
\vspace{0.2cm}
\centering 
\epsfig{file=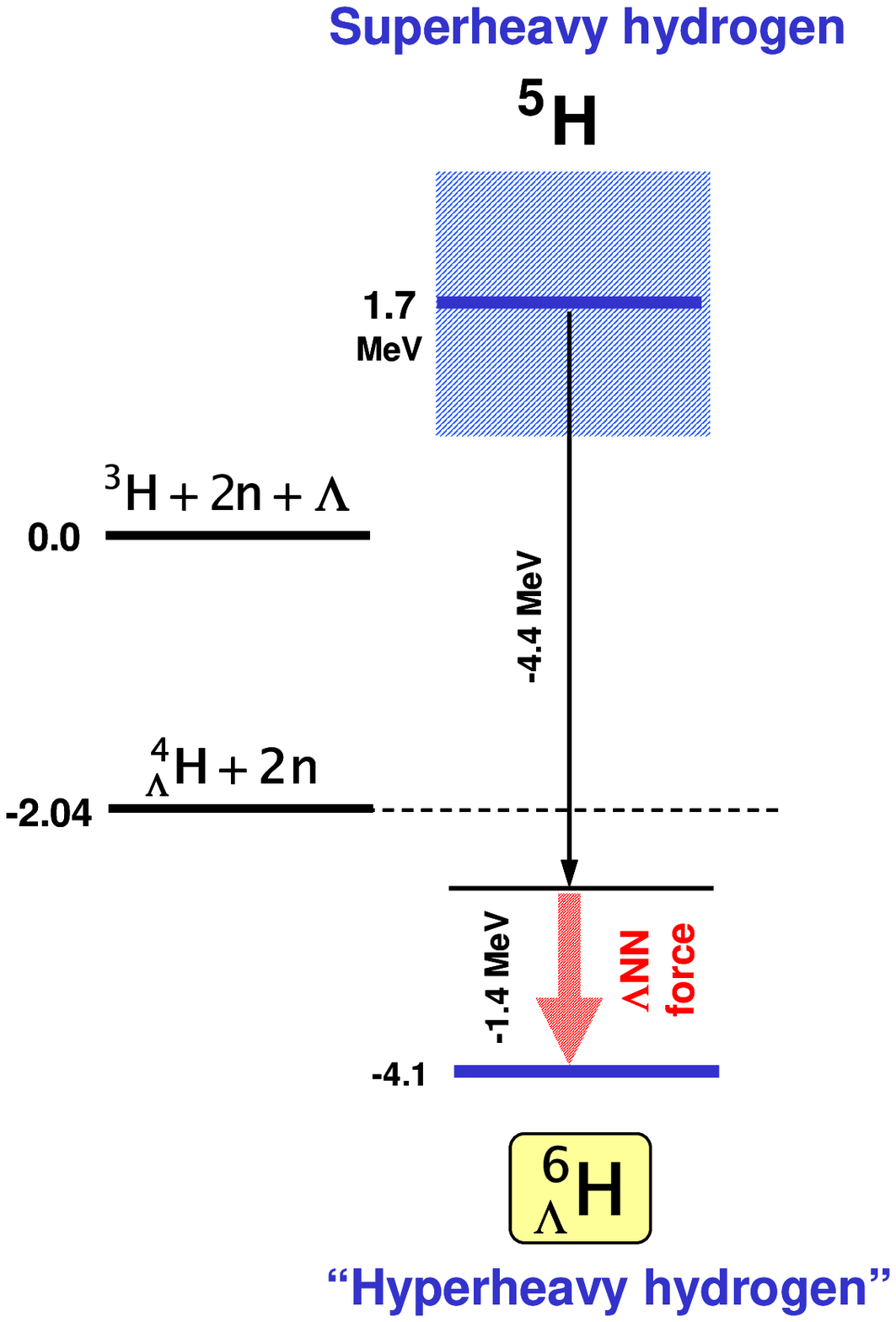, width=5cm}\\ 
\caption{\small \label{fig;pred}
Energy level of $^6_\Lambda$H$(0^+)$ predicted by M-A is shown together with the coherent $\Lambda$-$\Sigma$ coupling effect, i.e. the $\Lambda NN$ three-body force effect. }
\end{figure} 

A theoretical prediction of the "hyper-heavy" hydrogen based on realistic $\Lambda N$-$\Sigma N$ dynamics has been given by Myint and Akaishi (M-A) in 2002 \cite{Myint02,Shinmura02,Akaishi08} which is 10 years before the experimental observation of FINUDA. It is predicted to be a particle stable bound state with 2.1 MeV below the threshold of $^4_\Lambda$H$+2n$ as shown in Fig.~\ref{fig;pred}. 
When the coherent $\Lambda$-$\Sigma$ coupling term is turned off, the $\Lambda$-separation energy from $^5$H reduces to 4.4 MeV, and the state is only 0.7 MeV below the threshold. The $\Lambda N$-$\Sigma N$ coupling effects are coherently added to give 1.4 MeV attraction to the ground state. The coherent $\Lambda$-$\Sigma$ coupling plays a significant role in binding mechanism of this $^6_\Lambda$H system. It has been also shown that all the experimental binding energies of $s$-shell $\Lambda$-hypernuclei, $^3_\Lambda$H, $^4_\Lambda$H and $^5_\Lambda$He, are consistently reproduced by taking into consideration of the coherent $\Lambda$-$\Sigma$ coupling effect \cite{Akaishi00,Nemura02}.

\begin{figure}[htb!] 
\vspace{0.5cm}
\centering 
\epsfig{file=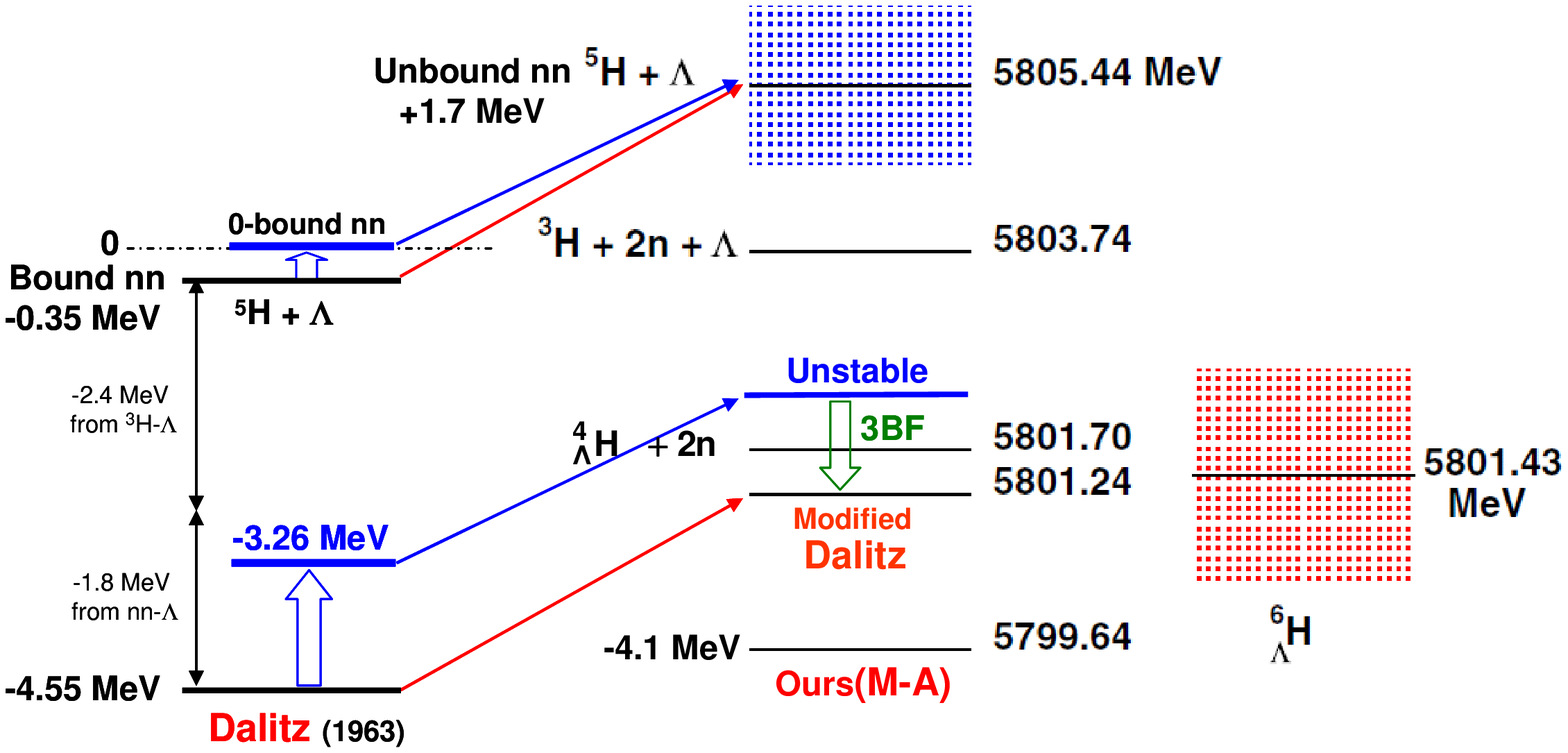, width=14cm}\\ 
\caption{\small \label{fig;shift}
Energy level shifts for $^6_\Lambda$H$(0^+)$ are shown together with the original and modified Dalitz's theoretical predictions and the FINUDA experimental value.}
\end{figure} 

\section{Theoretical analysis of $^6_\Lambda$H}

Experimental evidence of $^6_\Lambda$H has been given by Agnello et al. \cite{Agnello12,Agnello12a}: It is claimed that the experimental state is very close to Dalitz's theoretical prediction, and also claimed that our M-A result deviates from this experimental value. This statement, however, is based only on crude estimation. So, we make here a careful analysis on this matter. The binding energy of $^6_\Lambda$H was originally predicted by Dalitz in 1963 \cite{Dalitz63} to be 4.2 MeV from the old $^5$H+$\Lambda$ threshold level, which is similar to our result as shown in Fig.~\ref{fig;shift}. Later, the $^5$H was discovered to be a resonance state of +1.7 MeV rather than a weakly bound state of -0.35 MeV \cite{Ajzenberg57} from $^3$H$+2n$. The energy level referred to as Dalitz in \cite{Agnello12,Agnello12a} is 2 MeV shifted upwards by just adding the difference between the resonance and the bound states. Thus, this modified Dalitz state becomes very close to the experimental one. But we do not think, from dynamical point of view, that this shift due to bound $nn \rightarrow$ unbound $nn$ is reasonable. We have made a three-body calculation of  $t$-$(nn)$-$\Lambda$ with phenomenological potentials to reproduce Dalitz's original result. The phenomenological potentials are as follows,
\begin{eqnarray}
V_{t(nn)}(r) = -13.3~{\rm MeV}~~{\rm exp} [-(r/2.2~{\rm fm})^2], \label{eq;tnn} \\
V_{\Lambda t}(r) = -45.4~{\rm MeV}~~{\rm exp} [-(r/1.53~{\rm fm})^2], \label{eq;Lt} \\
V_{(nn)\Lambda}(r) = -11.5~{\rm MeV}~~{\rm exp} [-(r/1.8~{\rm fm})^2]. \label{eq;nnL} 
\end{eqnarray}
The binding energies of $^5$H and $^4_\Lambda$H systems obtained with the above first two interactions are 0.35 MeV ($t$-$nn$) and 2.4 MeV ($\Lambda$-$t$), respectively. The last one has no bound state, and is determined so as to reproduce the original Dalitz's binding, 4.55 MeV, of $t$-$(nn)$-$\Lambda$. Then, the interactions of Eqs. (\ref{eq;tnn}), (\ref{eq;Lt}) are modified to give the zero $(nn)$ binding from $t$ in $^5$H as a step to the resonance and to fit the presently accepted 2.04 MeV binding of $^4_\Lambda$H. The obtained interactions are as follows; 
\begin{eqnarray}
V_{t(nn)}(r) = -10.5~{\rm MeV}~~{\rm exp} [-(r/2.2~{\rm fm})^2], \label{eq;tnn'} \\
V_{\Lambda t}(r) = -43.8~{\rm MeV}~~{\rm exp} [-(r/1.53~{\rm fm})^2]. \label{eq;Lt'} 
\end{eqnarray}
As a result of these changes Dalitz's value of $^6_\Lambda$H binding is reduced from 4.55 MeV to 3.26 MeV measured from the zero level of $t$+$2n$+$\Lambda$. If we make a further change from the zero $(nn)$-binding to the 1.7 MeV resonance, the original Dalitz's state for $^6_\Lambda$H is shifted to be surely an unstable state above the $^4_\Lambda$H$+2n$ threshold, which cannot survive till its weak decay to emit $\pi^-$, as is shown in Fig.~\ref{fig;shift}. So some extra attraction like coherent $\Lambda NN$ three-body force is required to explain the FINUDA experimental data. Thus, the phenomenological shift to get the "modified Dalitz" done in \cite{Agnello12,Agnello12a} is too easygoing way, which is not justified from the dynamical point of view.

\begin{figure}[htb!] 
\vspace{0.5cm}
\centering 
\epsfig{file=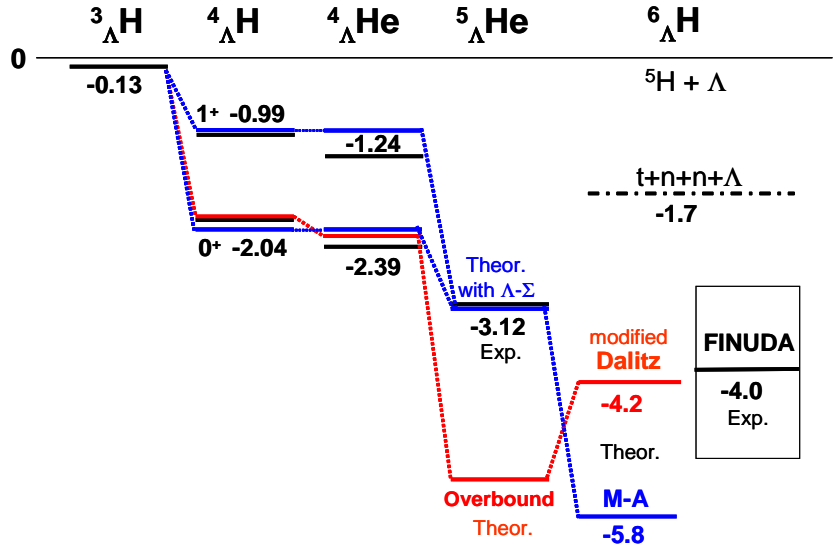, width=11cm}\\ 
\caption{\small \label{fig;Lsep}
Theoretical $\Lambda$ energy levels ($\Lambda$ separation energies) for $s$-shell hypernuclei and $p$-shell $^6_\Lambda$H are denoted together with respective experimental data (unit in MeV).}
\end{figure}

Our theoretical prediction (M-A) of $^6_\Lambda$H \cite{Myint02} is based on systematic analyses of three, four and five-body $s$-shell $\Lambda$-hypernuclei. All the empirical $\Lambda$-separation energies are consistently reproduced with realistic Nijmegen $YN$ potentials \cite{Rijken99}, as is shown in Fig.~\ref{fig;Lsep}.

It should be stressed that the long-standing "$^5_\Lambda$He overbinding problem" \cite{Dalitz72} has been solved in our treatment by taking into account the coherent $\Lambda$-$\Sigma$ coupling effect: only in the six-body $^6_\Lambda$H case, there is a deviation from the FINUDA data. Now we are going to check whether this deviation is serious or not. On the other hand, Dalitz's approach has to solve the most difficult problem of the $^5_\Lambda$He overbinding along with \cite{Dalitz72}. 

\begin{figure}[htb!] 
\vspace{0.5cm}
\centering 
\epsfig{file=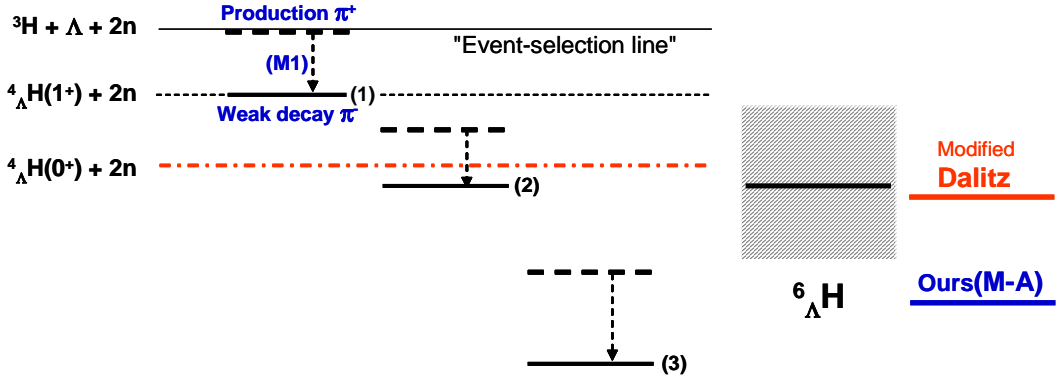, width=14cm}\\ 
\caption{\small \label{fig;event}
Three candidate pair-events for the neutron-rich hypernucleus, $^6_\Lambda$H \cite{Agnello12,Agnello12a}. The levels (1), (2), (3) detected by weak decay $\pi^-$ are assigned to be the ground state which can be compared safely with theoretical predictions. }
\end{figure}

The FINUDA experimental value of $^6_\Lambda$H is obtained from an average of three pair-events shown in Fig.~\ref{fig;event}. Among the levels (1), (2), (3) detected by $\pi^-$ it is quite unlikely from theoretical point of view to observe the event (1) as a weak decay of the ground state, $^6_\Lambda$H$(0^+)$, since the state lies above the strong-decay threshold, $^4_\Lambda$H$(0^+)+2n$. If we omit this event and take the average of the remaining two events, (2) and (3), the experimental result comes close to our M-A. We should wait for further experimental data with more statistics to conclude anything. 

\section{Underlying physics of $^6_\Lambda$H}

\begin{figure}[htb!] 
\vspace{0.5cm}\centering 
\epsfig{file=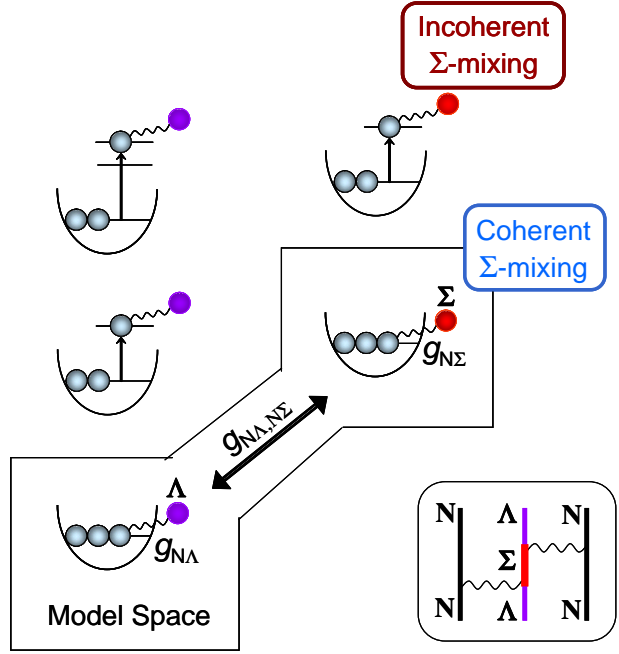, width=8cm}\\ 
\caption{\small \label{fig;coh}
Contributions from the coherent and from the incoherent $\Lambda$-$\Sigma$ couplings in $^4_\Lambda$H.}
\end{figure}

\begin{figure}[htb!] 
\vspace{0.5cm}\centering 
\epsfig{file=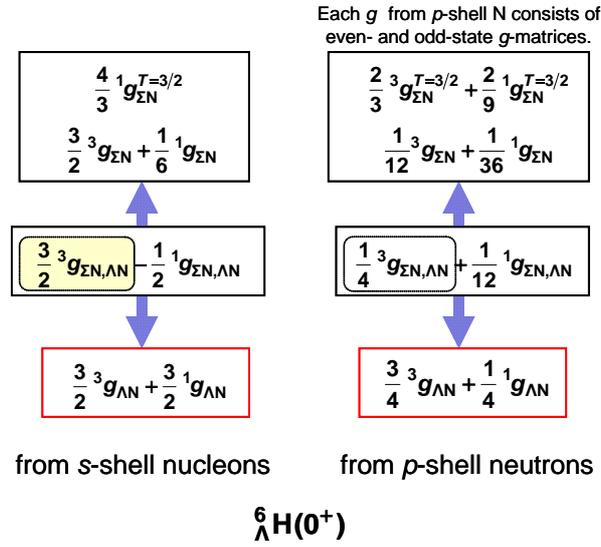, width=8cm}\\ 
\caption{\small \label{fig;coh2}
Coupled channel $YN$ interactions with spin-isospin weights for $^6_\Lambda$H$(0^+)$.}
\end{figure}

Theoretical fitting to data is just a starting point but not the goal of study for underlying physics: a naive comparison, done in \cite{Agnello12,Agnello12a}, of figures from phenomenological (Dalitz) and dynamical (M-A) theories at different stages is rather meaningless. In M-A, we have discussed the importance of coherent $\Lambda$-$\Sigma$ coupling effects in $^6_\Lambda$H as well as in $s$-shell $\Lambda$-hypernuclei. In Fig.~\ref{fig;coh} a process, where a nucleon changes to an excited state after the interaction, is called incoherent coupling. The coherent $\Lambda$-$\Sigma$ coupling is a process in which a nucleon remains in its ground configuration after converting $\Lambda$ to $\Sigma$, while giving all other nucleons an equal footing to interact with the $\Sigma$. The coherent $\Lambda$-$\Sigma$ coupling of $^6_\Lambda$H is obtained by folding a sum of interactions from three $s$-shell nucleons and from two $p$-shell neutrons which are schematically described in Fig.~\ref{fig;coh2}.

\begin{figure}[htb!] 
\vspace{0.5cm}
\centering 
\epsfig{file=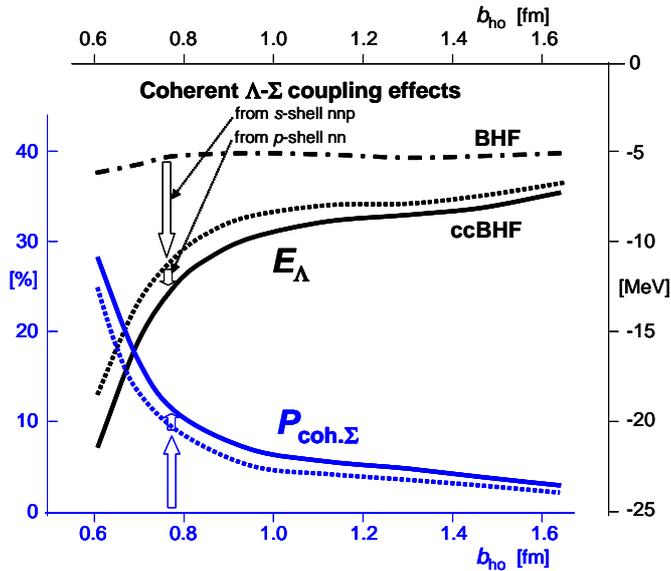, width=9cm}\\ 
\caption{\small \label{fig;size} 
Usual and extended BHF calculations done with D2 interaction \cite{Akaishi00} for the $^6_\Lambda$H system. The $\Lambda$ binding energy, $E_\Lambda$, and the coherent $\Sigma$ component, $P_{{\rm coh.} \Sigma}$, are given as functions of the core-nucleus size varied artificially by using a harmonic oscillator model, $\hbar \omega = \hbar^2/(M b_{\rm ho}^2)$. }
\end{figure}

In our studying of $^6_\Lambda$H we used an extended Brueckner-Hartree-Fock (BHF) method with coupled $\Lambda$-$\Sigma$ mean-fields \cite{Myint02,Akaishi08,Akaishi00} obtained from the NSC97f potential \cite{Rijken99}. The $\Lambda$-$\Sigma$ coupling effect was found to be 1.4 MeV with 1.6\% of the coherent $\Sigma$ component. The coherent $\Lambda$-$\Sigma$ coupling contribution from two $p$-shell neutrons in $^6_\Lambda$H is not so large due to dilute distribution of the two neutrons. It does not mean, however, non-significance of the coherent $\Lambda$-$\Sigma$ coupling in dense neutron-rich matter. Figure \ref{fig;size} demonstrates the dependence of $\Lambda$ binding energy, $E_\Lambda$, and coherent $\Sigma$ component, $P_{{\rm coh.}\Sigma}$, on the core-nucleus size of $^6_\Lambda$H. As the size becomes compact, the coherent effect grows to an appreciable amount in this neutron-rich system. 

It is noted that such coherent $\Lambda$-$\Sigma$ coupling effects disappear in symmetric $(N$=$Z)$ nuclear matter. Thus, in order to treat neutron-rich $\Lambda$-hypernuclei, the usual BHF method for symmetric case should be extended to the coupled-channel BHF method with the $\Lambda$-$\Sigma$ transition mean-field which causes the $\Lambda$-$\Sigma$ hyperon-mixing \cite{Shinmura02}.

\section{Concluding remarks}

The FINUDA data of $^6_\Lambda$H \cite{Agnello12,Agnello12a} could be a milestone towards the coherent $\Lambda$-$\Sigma$ hyperon-mixing in neutron-star matter \cite{Shinmura02}. In order to establish the strength of the coherent $\Lambda$-$\Sigma$ coupling, we need more experimental data with good statistics and more careful theoretical investigations of few-body neutron-rich hypernuclei. We have a plan to conduct thorough investigation on $^6_\Lambda$H as well as $^{10}_\Lambda$Li \cite{Saha05} within the coupled-channel Brueckner-Hartree-Fock framework. Concerning the heavy hyperhydrogen, $^6_\Lambda$H, some data from the J-PARC $(\pi^-,K^+)$ experiment \cite{Sakaguchi12} are highly awaited.

\section*{Acknowledgments}

One of the authors (Theingi) is grateful to the organizing committees of SNP12 and FB20 for supporting to attend the Workshop.

\end{document}